\documentclass[preprint2]{aastex}

\begin{document}
\title{Serendipitous Detections of XTE J1906+09 with the {\it Rossi
X-ray Timing Explorer}}

\author{Colleen A. Wilson\altaffilmark{1}, Mark H. Finger\altaffilmark{2}, Ersin
{G\"o\u{g}\"u\c{s}}\altaffilmark{3}, Peter M. Woods\altaffilmark{2}, Chryssa
Kouveliotou\altaffilmark{1,2}}
\affil{SD 50 Space Science Research Center, National Space Science and Technology
Center, 320 Sparkman Drive, Huntsville, AL 35805}
\email{colleen.wilson-hodge@msfc.nasa.gov}
\altaffiltext{1}{NASA's Marshall Space Flight Center}
\altaffiltext{2}{Universities Space Research Association}
\altaffiltext{3}{University of Alabama in Huntsville}

\begin{abstract}

In 1996 during {\em Rossi X-ray Timing Explorer (RXTE)} observations of SGR 1900+14, 
the 89-s X-ray pulsar XTE J1906+09 was discovered. As a result of
monitoring campaigns of SGR 1900+14, XTE J1906+09 was also monitored regularly
in 1996 September, 1998 May-June, 1998 August-1999 July, and 2000 March-2001
January. A search for pulsations in these observations resulted in detections of
only the two previously reported outbursts in 1996 September and 1998 August-
September. Pulsed flux upper limits for the remaining observations indicate that
XTE J1906+09 is a transient X-ray pulsar and therefore likely has a Be star 
companion. XTE J1906+09 was not detected with the {\em RXTE} All Sky Monitor. Pulse 
timing analysis of the second outburst revealed a sinusoidal
signature in the pulse frequencies that is likely produced by periastron passage
in an orbit. Fits to pulse phases using an orbital model and quadratic
phase model have $\chi^2$ minima at orbital periods of 26-30 days for fixed
mass functions of 5, 10, 15, and 20 $M_{\odot}$. The pulse shape showed 
intensity and energy dependent variations. Pulse phase
spectroscopy was used to quantify the energy dependent variations. The phase 
averaged spectrum, using the pulse minimum spectrum as the background spectrum 
to eliminate effects from SGR 1900+14 and the galactic ridge, was well fitted 
by an absorbed power law with a high energy cutoff with column density 
$N_{\rm H} = (6 \pm 1) \times 10^{22}$ cm$^{-2}$, photon index = 
$1.01 \pm 0.08$, cutoff energy $E_{\rm cut} = 11 \pm 1$ keV, and folding energy
$E_{\rm fold} = 19 \pm 4$ keV. Estimated 2-10 keV peak fluxes, corrected for 
contributions from the galactic ridge and SGR 1900+14, are $6 \times 10^{-12}$ ergs cm$^{-2}$ s$^{-1}$ and 
$1.1 \times 10^{-10}$ ergs cm$^{-2}$ s$^{-1}$ for the 1996 and 1998 
outbursts, respectively. XTE J1906+09 may be a member of
an unusual class of Be/X-ray binaries that do not lie on the general spin period
versus orbital period correlation with the majority of Be/X-ray binaries.
\end{abstract}


\section{Introduction}

XTE J1906+09 is an 89-s accreting X-ray pulsar that was serendipitously
discovered in a 1996 {\em Rossi X-ray Timing Explorer (RXTE)} Proportional Counter
Array (PCA) observation of SGR 1900+14 \citep{Marsden98}. Two outbursts from
XTE J1906+09 have been reported to date, in 1996 September \citep{Marsden98}
and 1998 August-September \citep{Takeshima98}. The 1996 energy spectrum was well
fitted with either a  thermal Bremsstrahlung spectral shape with temperature
$19.5 \pm 4.6$ keV and $N_{\rm H} = (7.7 \pm 5.2) \times 10^{22}$ cm$^{-2}$ or
a power law with a photon index $1.9 \pm 0.1$ and $N_{\rm H} = (10.8 \pm 4.0)
\times 10^{22}$ cm$^{-2}$\citep{Marsden98}, when a fixed spectrum was assumed for the galactic
ridge emission \citep{Valinia98}. The high absorption column density suggested 
that this source is at a distance of $\gtrsim 10$ kpc. \citet{Marsden98}  
suggest that XTE J1906+09 could have either a Be star companion or a supergiant
companion that is under-filling its Roche lobe.

To date, about 80 accretion powered pulsars are known. These systems are often 
divided into 2 classes based on the mass of their companion: (1) high mass X-ray 
binaries (HMXBs) with OB companions (75 of the known systems) and (2) low mass 
X-ray binaries (5 of the known systems) with type A (or later) companions. The 
HMXBs are usually further divided into 3 classes based on the
accretion mode of the system: (1) pulsars accreting from Roche lobe filling
supergiants that tend to have short spin and orbital periods, (2) pulsars
accreting from the strong stellar wind of a supergiant companion which is 
under-filling its Roche lobe that tend to have longer orbital periods and spin
periods and show persistent but highly variable emission, and (3) pulsars
accreting from the circumstellar disk around a Be star that as a population show
a correlation between their spin and orbital periods
\citep{Corbet86, Waters89}. 

Be/X-ray binaries are the most common HMXBs. They consist of a pulsar and a Be
(or Oe) star, a main  sequence star of spectral type B (or O) that shows Balmer
emission lines (See e.g., Slettebak 1988 for a review.) The line emission is 
believed to be associated with circumstellar material shed by the Be star into 
its equatorial plane. The exact nature of the mass loss process is unknown, but
it is thought to be related to the rapid rotation of the Be star, typically near
70\% of the critical break-up velocity \citep{Porter96}. The equatorial material
forms a slow, dense outflow, which is generally believed to fuel the X-ray 
outbursts. Near the Be star, the equatorial outflow probably forms a 
quasi-Keplerian disk \citep{Quirrenbach97,Hanuschik96}. X-ray outbursts are 
produced when the pulsar interacts with the Be star's disk. Be/X-ray binaries \
typically show two types of outburst behavior: (i) giant outbursts, 
characterized by high luminosities and high spin-up rates (i.e., a significant 
increase in pulse frequency) and (ii) normal outbursts, characterized by lower 
luminosities, low spin-up rates (if any), and repeated occurrence at the orbital
period \citep{Stella86, Bildsten97}.

The target of the {\em RXTE} observations, SGR 1900+14, is a member of a small 
class (4 objects), called Soft Gamma Repeaters (SGR), which exhibit brief 
($\sim 0.1$ s), intense ($10^3-10^4 L_{\rm Edd}$) bursts of low energy gamma rays with
recurrence times ranging from seconds to years (See Hurley 2000 for a
recent review.) These objects also emit X-rays in quiescence. A spin period of
5.16-s was
discovered in the quiescent emission of SGR 1900+14 \citep{Hurley99}. 
The soft X-ray spectrum (0.1-10 keV) of SGR 1900+14 consists of a blackbody (kT
$\sim 0.5$ keV) plus a power law component ($\Gamma \sim -2$). The nominal
unabsorbed 2-10 keV flux level is $\sim 10^{-11}$ ergs cm$^{-2}$ s$^{-1}$.
However, both the flux and spectrum of SGR 1900+14 have been shown to vary
substantially during periods of burst activity \citep{Woods01}. Effects from SGR 1900+14 were
removed from the data whenever possible, but since SGR 1900+14 is always present
in the field of view when XTE J1906+09 is active, it contaminates measurements
of the total flux and average spectrum of XTE J1906+09.

In this paper we will present results of a search for additional outbursts of 
XTE J1906+09 in {\em RXTE} PCA observations from 1996-2001, detailed timing and
spectral analysis of the 1998 outburst, a comparison of the 1998 and 1996 
outbursts, and a discussion of the possible nature of this system.

\section{Observations}

The results presented in this paper involve data from the Proportional-Counter 
Array (PCA, Jahoda et al.\ 1996) and the High-Energy X-ray Timing Experiment 
(HEXTE, Rothschild et al.\ 1998) on the {\em Rossi X-ray Timing Experiment
(RXTE}, Bradt, Rothschild, \& Swank 1993). XTE J1906+09 is too weak to be detected with the {\em RXTE} All-Sky
Monitor (ASM, Levine et al.\ 1996). The PCA consists of five identical 
xenon/methane multi-anode proportional counter units (PCU) sensitive to photons
from 2-60 keV and has a total collecting area of 6500 cm$^2$. All 5 of the PCU 
are not always on, so in this paper, we have scaled all of our count rates to the rate that would have been
observed in 5 PCU. Both the PCA and HEXTE are collimated instruments with 
an approximately circular field-of-view with a FWHM of about 1\arcdeg\ 
\citep{Jahoda96, Rothschild98}. HEXTE consists of two clusters each containing 
four NaI/CsI scintillation detectors sensitive to photons in the 15-250 keV 
range. Each cluster has a collecting area of 800 cm$^{-2}$. One of the four
detectors in cluster B has lost all energy resolution, so this detector is not
included in our analysis. Each cluster can rock along mutually orthogonal
directions to provide background measurements 1.5 to 3.0\arcdeg\ away from the
measured source every 16-128 s \citep{Rothschild98}.

Because XTE J1906+09 is located 33\arcmin\ from  
SGR 1900+14, a large number of PCA and HEXTE observations 
have been made of this region. Table~\ref{tab:obs} lists the {\em RXTE} Cycle, 
proposal identification number, number of observations, total on source time in
kiloseconds, and the date ranges of the observations. Table~\ref{tab:pcus} lists
the percentage of time 5, 4, 3, or 2 PCUs were active for each row in
Table~\ref{tab:obs}.

\section{Timing Analysis}

\subsection{Frequency Search}

\begin{figure}
\plotone{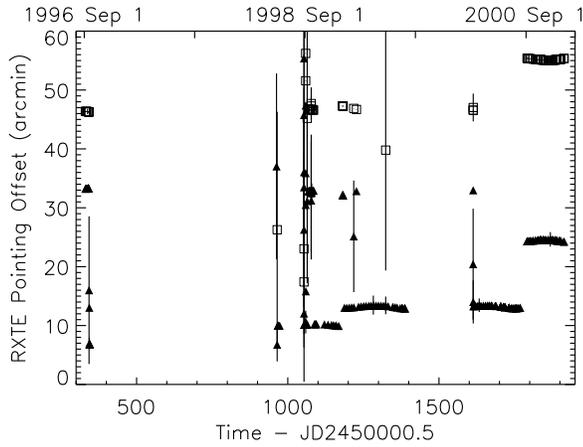}
\caption{{\em RXTE} pointing offsets to XTE J1906+09 (filled triangles) and 4U
1907+097 (open squares) during SGR 1900+14 observations. Error bars indicate 
the standard deviation in the pointing offset.  Scanning observations have the 
largest error bars. Gaps indicate times when each object was outside the 
{\em RXTE} field of view.
\label{fig:offset}}
\end{figure}

Data from each observation were analyzed using
FTOOLS\footnote{\url{http://heasarc.gsfc.nasa.gov/ftools}} 5.0
\citep{Blackburn95}. An {\em RXTE} filter file was made
for each observation. Good time intervals were selected where XTE J1906+09 was
in the {\em RXTE} field-of-view, was not Earth occulted, and the time since the
last SAA passage was greater than 30 s. Light curves with 16-s time resolution 
were produced selecting 2-30 keV Standard2 data from the top layer only of the 
PCA to improve signal to noise. Corresponding background light curves were 
produced using the FTOOL {\it pcabackest} with the faint source models. Net 
light curves were then produced by subtracting the background light curves from
the Standard2 light curves. Variances on the net count rates were computed as 
$\sigma_{\rm net}^2 = \sigma^2_{\rm total}+(0.01R_{\rm back})^2$ where 
$R_{\rm back}$ is the model background rate. The background model has an 
estimated systematic error of 1\% \citep{Jahoda96}. Because these observations 
are optimized for SGR  1900+14, not XTE J1906+09, the PCA pointing direction was
frequently offset significantly from the XTE J1906+09 position. Pointing 
offsets to XTE J1906+09 (filled triangles) and the bright 440-s pulsar 4U 
1907+09 (open squares) are shown in Figure~\ref{fig:offset}.  Error bars denote the
standard deviation in the pointing offset. Scanning observations have the
largest error bars. Gaps indicate times when the PCA did not observe this 
region of the sky. The  FTOOL {\it xtecol} was used to create collimator 
corrected light curves for the offset pointing. These light curves were then 
barycentered to the solar system barycenter using the FTOOLs {\it fxbary} 
(before 2000 December 31) and {\it faxbary} (after 2000 December 31). Bursts 
from SGR 1900+14, present in observations from 1998 August-September, were 
removed from the data using a simple algorithm that discarded data outliers. 
Outliers were defined as points where $|R-R_{\rm med}|>R_{\rm med}$ where $R$ is
the background subtracted count rate (in 16-s intervals) and $R_{\rm med}$ is 
the median background subtracted count rate (in 16-s intervals) for that 
observation.

A search for pulsations from XTE J1906+09 was performed for each
observation. Observations separated by less than 4 {\em RXTE} orbits were combined and
count rates for all observations were corrected for varying numbers of PCUs.
The data from each observation were fitted with a model consisting of a constant
term plus a second order Fourier expansion in a pulse phase model. The pulse
phase model used was of the form $\phi=\nu(t-t_{\rm mid})$ where $\nu$ is a
trial frequency, $t$ is the barycentered time associated with the net count
rate, and $t_{\rm mid}$ is the midpoint time of the observation. The trial frequency
search grid consisted of 81 evenly spaced frequencies from 11.168 to 11.261 mHz.
To correct the variances on the harmonic\footnote{In this paper, harmonics are defined as
$n\nu$ where $n = 1,2,3, \ldots$ and $\nu$ is the pulse frequency.}  amplitudes in the Fourier expansion for
aperiodic noise in the power spectrum (due either to XTE J1906+09 or other 
sources in the {\em RXTE} field of view), we first computed the average Leahy 
normalized \citep{Leahy83} power spectrum for 1 ksec segments of data and subtracted the 
pulsed component. Then we computed the average power $\bar P_n$ for the 
frequency intervals $[n\nu_0/2, 3n\nu_0/2]$, where $\nu_0$ is the pulse frequency
and $n=1,2, \ldots$ is the harmonic number. If $\bar P_n$ was greater than 
the expected Poisson level of 2.0, we multiplied the variances on the 
corresponding harmonic amplitudes by $\bar P_n/2$ to correct for aperiodic noise. This
correction accounts for aperiodic noise dependent systematic differences in 
sensitivity from observation to observation. The best fit frequency for each
observation was determined using a modified $Z^2_2$ statistic \citep{Buccheri83},
which we will call $Y_n$, that incorporates the corrected variances 
\citep{Finger99}.  For the best fit frequency, the root-mean-squared (rms) pulsed
amplitude was also computed. Results of the XTE J1906+09 frequency search are 
shown in Figure~\ref{fig:freq1906}. Monte Carlo
simulations were run to determine detection confidence levels. From $10^5$
trials using the time structure of the observation on 1999 January 9 (a typical
observation lasting about 15 ksecs with 10 ksecs on source) we 
determined detection confidence levels of 99\% ($Y_2 \gtrsim 14.6$), 99.9\% ($Y_2 \gtrsim 20.9$), and 
99.99\% ($Y_2 \gtrsim 29.0$). These confidence intervals do not take into
account effects from 4U 1907+09. 

The bright 440-s X-ray pulsar 4U 1907+09 \citep{IntZand98} is located 
near both XTE J1906+09 and SGR 1900+14 and was present in the PCA field of view
for some of the observations. Before performing a search for pulsations from 4U
1907+09, the net 2-30 keV light 
curves were corrected for the PCA collimator response to the offset pointing to
4U 1907+09 and were barycentered for its position. To test whether or
not this source was active, we performed a coarse frequency search over a grid 
of 31 evenly spaced frequencies from 2.225 to 2.341 mHz using the observations
where 4U 1907+09 was in the PCA field of view. By chance, the fifth
harmonic of 4U 1907+09's 440-s period is about 88-s, quite close to the 
89-s period of XTE J1906+09, so only the first 4 harmonics of 4U 1907+09
were used in the frequency search. The two sources can usually be easily 
separated in power spectra and were successfully separated in all observations
used for subsequent pulse phase analysis. The frequency search for 4U 1907+09 
told us when that source was active and when we needed to check for 
contamination in XTE J1906+09 measurements. Results of the 4U 1907+09 frequency
search are shown in Figure~\ref{fig:freq4u1907}. From $10^5$ Monte Carlo trials using the time 
structure of the observation on 1999 January 5 (a typical observation lasting
about 15 ksecs with 10 ksecs of on source time), we determined detection 
confidence levels of 99\% ($Y_4 \gtrsim 26.2$), 99.9\% ($Y_4 \gtrsim 39.3$), and 
99.99\% ($Y_4 \gtrsim 47.3$). 

\subsection{Pulse Phase Analysis}

To generate pulse phase measurements, we first made 2-30 keV light curves with 
1-s time resolution for 15 PCA observations (MJD 50337-50346, 1996 September 11-20 
in science data mode E\_125us\_64M\_0\_1s) during the first outburst and 28 PCA
observations (MJD 51056-51086, 1998 August 31-September 30, 13 in science data 
mode E\_125us\_64M\_0\_1s, 15 in science data 
mode Good Xenon) during the second outburst. The observations used for pulse
phase analysis included all those with significant frequency detections in our
grid search plus a few observations either side of those intervals of frequency
detections. The data were background subtracted
using a spline fit to the 16-s background model count rates generated with
the FTOOL {\it pcabackest} and errors on the net count rates were computed
assuming a 1\% systematic error on the background. Collimator corrections
generated for Standard 2 data using the FTOOL {\it xtecol} were applied 
manually. For each observation within each outburst, initial pulse profiles were
created by fitting harmonic
amplitudes in a Fourier series expansion in a phase model of the form 
\begin{equation}
\phi = \bar\nu(t-t_{\rm mid}) \label{eqn:phimod0}
\end{equation}
where $\bar\nu$ is the average pulse frequency for each outburst (11.2150 mHz and
11.2158 mHz for the first and second outbursts, respectively) from detections
in our grid search and $t_{\rm mid}$ is the mid-time of each outburst. Template
pulse profiles were 
also created by fitting harmonic amplitudes in a Fourier expansion in a phase
model of the form in Equation~\ref{eqn:phimod0}. For the first outburst, we used
the brightest interval 50343.533-50343.838 (1996 September 17) to generate the
template profile, with a pulse frequency of 11.216 mHz from our grid search.
For the second outburst, we used a moderate intensity interval MJD 
51072.829-51072.989 (1998 September 16) to create the template, with a pulse
frequency of 11.213 mHz from our grid search results. Figure~\ref{fig:tamp} 
(top panel) shows the pulse profile used to generate the template for the 
first outburst. Only the first 6 harmonics were significant. The bottom panel 
shows the pulse profile used to
create the template for the second outburst. Only the first 12 harmonics were
significant.  The mean has been subtracted from these pulse profiles and for all
other pulse profiles in this paper because the mean count rate includes
contributions from XTE J1906+09, SGR 1900+14, and in some observations, 4U
1907+097. Since RXTE is not an imaging instrument, the relative contribution 
from each of the three sources cannot be determined. Template pulse profiles
were 
normalized so that the sum of the squares of the harmonic amplitudes was equal 
to 2. Phase offsets and pulsed intensities were calculated by cross-correlating the fitted profile for each 
observation with the template profile. In the cross-correlations, 6 
harmonics were used for the first outburst and 12 harmonics for the second
outburst. Using the measured pulse frequencies from
our frequency search, we determined that we could phase connect
data from only 8 observations in the first outburst (MJD 50340.9-50344.8, 1996
September 14-19) and 23 observations from the second outburst (MJD 51056.44-
51075.64, 1998 August 31-September 20). We fitted the connected phase 
measurements from each outburst with a polynomial model of the form
\begin{equation}
\phi = \sum^n_{i=0} c_i(t-t_0)^i
\end{equation}
where $c_i$ are coefficients given in Table~\ref{tab:poly} and $t_0$ is an
epoch. Residuals to these polynomial models are shown in Figure~\ref{fig:resid}. 
Pulse frequencies computed by differencing adjacent connected phases show a 
period of spin-up followed by spin-down followed by spin-up, shown in 
Figure~\ref{fig:outburst2} (top panel). This is strongly suggestive of
an orbital signature and is consistent with either a short, nearly circular 
orbit or with periastron passage in a much longer eccentric orbit. No clear
signature is seen in measurements from the first outburst; however, this is not
surprising since the phase connected measurements only span 4 days and the
outburst was much weaker.

To investigate possible orbits, we fitted a quadratic phase model and an orbit 
with a fixed period and fixed mass function to 23 connected phase measurements 
from the second outburst. The connected phase measurements from the first 
outburst spanned too short a time to improve our fits. A grid of 300 orbital 
periods from 20 to 320 days was searched for fixed mass functions of $f(M) = $ 5,
10, 15, 20 $M_{\odot}$. (Be stars typically have masses in the range 8-20 
$M_{\odot}$.) Figure~\ref{fig:porb_grid} shows $\chi^2$ versus orbital
period for the selected mass functions. For all four mass functions, the best 
fit period was in the range 26-30 days, the eccentricity was low, 0.03-0.06, and
the frequency derivative was relatively large $1-2 \times 10^{-11}$ Hz s$^{-1}$.
The periastron epoch was in the range JD2451055-57 and the periapse angle was in
the range 126-144\arcdeg.  We advise caution using these parameters because this
apparent orbital signature was observed in a single 25-day interval.
Observations of additional outbursts are needed to confirm these parameters.

\subsection{Pulse Profiles}

The 2-30 keV pulse profiles shown in Figure~\ref{fig:tamp} are similar
despite being from very different time intervals and intensities. Both consist 
of 2 main peaks with the first peak (phase $\sim 0.1-0.4$ top panel) being 
narrower than the second peak (phase $\sim 0.5-1.1$ top panel). During the 1996
September outburst, the source was much fainter than during the 1998 outburst.
Comparing the two profiles suggests that a correlation between the pulse shape 
and intensity may be present. The first main peak is much brighter relative to 
the second main peak in the pulse profile from the stronger outburst than in 
the weaker outburst. To further investigate this effect, we phase aligned data 
from the 1998 outburst using the phase model in Table~\ref{tab:poly}. The phase
aligned harmonic amplitudes were averaged over 5 bands of 2-30 keV rms pulsed
flux: 0-5, 5-10, 10-15, 15-20, and 20-25 counts sec$^{-1}$. 
Figure~\ref{fig:profile_intensity} shows the average profiles from the 5 
intensity bands (top 5 panels) and the profile from the brightest interval in
the 1996 outburst (bottom panel and top panel of Figure~\ref{fig:tamp}). The 
intensity of the first main peak (phase $\sim 0.2-0.5$) increases dramatically 
as the pulsed flux increases. The second main peak (phase $\sim 0.6-1.2$) 
gradually evolves from a broad peak approximately equal in intensity to the 
first main peak to multiple peaks with considerable structure as the pulsed flux
increases. Within the 1998 outburst, the pulse shape appears to be correlated 
with intensity. The pulse profile from the 1996 outburst looks consistent with 
the low intensity profiles from the 1998 outburst, suggesting that the pulse 
shape-intensity correlation holds for both outbursts.

Next we extracted light curves in several energy bands to look for energy
dependence in the shape of the pulse profile. These light curves were background
subtracted, collimator corrected, and barycentered. For the first outburst, the
5 brightest observations when 4U 1907+09 was not present in the PCA field of
view (MJD 50343.6-50344.8, 1996 September 17-18) were combined and epoch-folded in the time domain 
using the phase model in Table~\ref{tab:poly} in 3 energy bands shown in 
Figure~\ref{fig:profile_energy1}. Corresponding background subtracted event mode
HEXTE light curves were extracted using the FTOOL {\it hxtltcurv}, barycentered,
and collimator corrected. \citet{Marsden98} report a detection of XTE
J1906+09 with HEXTE during this outburst, however our HEXTE pulse profiles are
consistent with a constant value, i.e., non-detection of XTE J1906+09. We are
not certain which observations \citet{Marsden98} used, since there were no {\em
RXTE} observations of this region on 1996 August 16-19, the dates reported in 
their paper. It appears that they also used the 1996 September observations
based on our comparisons with their PCA results. Figure~2 in \citet{Marsden98} 
shows folded pulse profiles from HEXTE plotted in units of counts bin$^{-1}$. 
If we fold our HEXTE data, also uncorrected for the total time in each bin, we 
see a similar profile. When we correct for the total time in each phase bin 
(i.e., plotting in units of counts s$^{-1}$), our profiles are consistent with 
a constant value. Further, the 20-30 keV HEXTE profile in \citet{Marsden98} does 
not appear to agree with the PCA profile in the same band, while in the 1998
outburst, where we clearly detect XTE J1906+09 with HEXTE and the PCA, the
profiles in this band agree well. 

For the second outburst, we selected the observation made on MJD 
51064 (1998 September 8) because it was near the peak of the outburst, it was a
long observation ($\sim 8300$ s of on-source time), and 4U 1907+09 was 
not present in the PCA field of view.  PCA light curves were extracted in 5 
energy bands, 2-5, 5-10, 10-15, 15-20, and 20-30 keV. HEXTE light curves were 
extracted in 3 energy bands 15-20 keV, 20-30 keV, and 30-100 keV. HEXTE was in 
``staring" mode for this observation, so no background measurements were 
available. These light curves were epoch-folded using the phase model in 
Table~\ref{tab:poly} and are shown in Figure~\ref{fig:profile_energy2}. The 
collimator correction was applied to the HEXTE pulse profiles by multiplying 
the epoch-folded mean-subtracted rates by the correction faction computed using
the FTOOL {\it xtecol}. At low energies ($<10$ keV), the pulse profile
consists of 3 peaks: a bright peak from phase $\sim 0.21-0.50$, a broad peak 
from phase $\sim 0.6-1.05$, and a narrow peak from phase $\sim 1.07-1.15$. As 
energy increases, the brightest peak remains distinct, the minimum following the
first peak becomes broader, and the second two peaks blend into a broad 
shoulder. 

\section{Spectral Analysis}

We selected the $\sim 8300$ s observation on MJD 51064 (1998 September 8) for 
extraction of a phase averaged spectrum of XTE J1906+09. This observation was 
selected because no bursts from SGR 1900+14 were present and 4U 1907+09 was not
in the {\em RXTE} field-of-view. As mentioned above, HEXTE was in staring mode 
for this observation, so HEXTE spectra were not used because background data 
were not available. PCA data had better signal to noise  for the 15-30 keV band,
where XTE J1906+09 was detected, than HEXTE.  Spectra  were
extracted from Good Xenon (256 channel, photon time tagged) PCA data for the
top Xenon layer only and corresponding PCA  background spectra were generated
using the faint source background model in  {\it pcabackest}. Response matrices
were created using the FTOOL {\it pcarsp}  with the XTE J1906+09 location. The
phase averaged PCA data were fitted using XSPEC 11.0.1. For  the spectral fits,
PCA data in the energy range 2.5-30 keV were used. The spectrum was best fitted
with a typical accreting X-ray  pulsar spectrum, an absorbed power law with a
high energy cutoff + a Gaussian  iron line. The best fit parameters are listed
in the second column of  Table~\ref{tab:spectra}. Evidence for an iron line at
$\sim 6.6$ keV is suggested by our spectral fits. The 2-10 keV flux computed
using this spectral model is $1.59 \times 10^{-10}$ ergs cm$^{-2}$ s$^{-1}$.
From Figure 2 in \citet{Woods01}, we estimated the 2-10 keV flux from SGR 
1900+14 as $3.2 \times 10^{-11}$ ergs cm$^{-2}$ s$^{-1}$. The spectral fit from
the 1996 outburst in \citet{Marsden98}, gave a 2-10 keV flux of $1.67 \times 
10^{-11}$ ergs cm$^{-2}$ s$^{-1}$ for the galactic ridge. Subtracting the 
contaminating fluxes, we estimated the 2-10 keV flux from XTE J1906+09 was 
$1.1 \times 10^{-10}$ erg cm$^{-2}$ s$^{-1}$. This flux is listed in 
Table~\ref{tab:spectra}.  Observations with an imaging instrument are needed to
fully separate the un-pulsed component of the XTE J1906+09 spectrum from the
galactic ridge and SGR 1900+14.

To investigate if the iron line was likely due to XTE J1906+09 and to look for
spectral evolution, we generated phase averaged spectra, background spectra, and
response matrices as described in the previous paragraph for 10 observations 
from 1998 August 31 - September 9 (MJD 51056-51065) using Standard 2 (129
channel, 16-s time resolution) PCA data for the top Xenon layer only. 
Observations after 1998 September 9 were not used because 4U 1907+09 was in the
PCA field of view. For each observation, PCA data in the energy range 2.5-30 keV
were fitted with an absorbed power law with a high energy cutoff + a
Gaussian iron line. As the intensity increased, the photon power law index 
decreased from $\Gamma \sim 2$ to $\Gamma \sim 1.2$. This apparent spectral 
hardening may be intrinsic to the source or it could be due to contamination 
from the softer spectra of SGR 1900+14 ($\Gamma \sim 2$, Woods et al.\ 2001) and
the galactic ridge ($\Gamma \sim 1.8$, Valinia \& Marshall 1998). Figure~\ref{fig:iron} shows the integrated flux of the iron 
line versus the rms pulsed flux for each observation. The iron line flux appears
to be correlated with the pulsed flux, with a correlation coefficient of 0.93
and a chance probability of $10^{-4}$; however, the errors on the iron line flux
are fairly large. This correlation suggests that the iron line is intrinsic to XTE J1906+09. 
 
To study the energy spectrum of the pulsed emission we created phase resolved
energy spectra for 20 equal phase intervals with the FTOOL {\it fasebin} using PCA 
good xenon data from MJD 51064 (1998 September 8). Instead of using the 
background model to generate the background spectrum, we used the measured 
spectrum at pulse minimum (phase 0.5-0.6) as the background spectrum 
in XSPEC. These spectra were well fitted with a typical pulsar model, an
absorbed power law with a high energy cutoff. No iron line was required to
obtain acceptable fits. The phase averaged spectrum was again fitted, but this
time using the pulse minimum as the background spectrum. The parameters of this
fit are listed in the third column of Table~\ref{tab:spectra}. 

Fit results for
the spectra from each of the 19 phase intervals using the pulse minimum spectrum
as the background are shown in Figure~\ref{fig:phaseresolved}. The fit results
from the phase averaged spectrum are plotted as dotted lines. Panel (a) shows
the mean-subtracted 2-30 keV count rate in each phase bin. Arrows indicate the 
bin used as the background spectrum. The absorption column density $N_H$, 
plotted in panel (b), increases for the bin centered on phase 0.15. This
corresponds to the dip at phase $\sim 0.15$ in
Figure~\ref{fig:profile_energy2}, just before the bright peak. This dip becomes
shallower with increasing energy and has disappeared above about 15 keV. Panels
(c), (d), and (e) show the power law photon index, the cutoff energy 
$E_{\rm cut}$, and the e-folding energy $E_{\rm fold}$, respectively. These 
panels show evidence for spectral softening around the pulse minimum in the 
phase range $\sim 0.4-0.6$. In Figure~\ref{fig:profile_energy2} the pulse
minimum becomes broader with increasing energy, also indicating spectral
softening. The hardest spectral index is at phase 0.15, corresponding to the dip
present at low energies at phase $\sim 0.15$, that disappears as energy
increases, suggesting spectral hardening. 

Using our spectral model, we computed a phase averaged 2-10 keV pulsed flux of 
$7.23 \times 10^{-11}$ ergs cm$^{-2}$ s$^{-1}$. Taking the ratio of the pulsed
flux to the corrected 2-10 keV total flux in Table~\ref{tab:spectra} gives a 
peak-to-peak pulse fraction of $\sim 66\%$. If we assume this pulse fraction is
constant for both outbursts, we estimate the total 2-10 keV flux in the 1996 
outburst was $6 \times 10^{-12}$ ergs cm$^{-2}$ s$^{-1}$ from our pulsed flux
measurements. The quiescent 2-10 keV flux from SGR 1900+14 is believed to be 
constant at a level of about $1 \times 10^{-11}$ ergs cm$^{-2}$ s$^{-1}$
\citep{Woods01}. Hence, the combined XTE J1906+09 and SGR 1900+14 would be 
$1.6 \times 10^{-11}$ ergs cm$^{-2}$ s$^{-1}$, consistent with the 2-10 keV flux
measured from both sources by \citet{Marsden98}. This suggests that the XTE
J1906+09 pulse fraction in the 2-10 keV band is approximately constant at 66\%.

\section{Discussion}
The sky region including XTE J1906+09 was regularly monitored with {\em RXTE} 
every few days from 1998 August to 1999 July and again from 2000 March - 2001 January as part 
of our guest investigation to observe SGR 1900+14. Prior to the monitoring
campaign, this sky region was observed regularly for about 2 weeks in 1996
September and again in 1998 May-June. A search for pulsations from XTE J1906+09 
in these observations resulted in the detection of two outbursts, 1996 September
17-18 (MJD 50343-50344) and 1998 August 30 - September 24 (MJD 51056-51080).  
Both outbursts have been previously reported \citep{Marsden98, Takeshima98}. 
Unfortunately our data cannot unambiguously determine the orbital period of the 
system. The two detected outbursts are separated by about 730 days. This spacing
may be an integer multiple of the orbital period, since normal outbursts in 
transient X-ray pulsars usually occur near periastron passage. Outbursts similar
to the 1996 outburst, which was detected for only 2 days in the frequency grid 
search and for about 4 days in pulse phase measurements, could have been easily
missed because the observations of the region were usually at least few days 
apart even during the regular monitoring. The non-detection of XTE J1906+09 in 
most of the observations tells us that this source is a transient source.
The known companions of transient X-ray pulsars are all Be stars (with the
exception of GRO J1744--28), hence XTE J1906+09 most likely has a Be star 
companion. 

\begin{figure}
\plotone{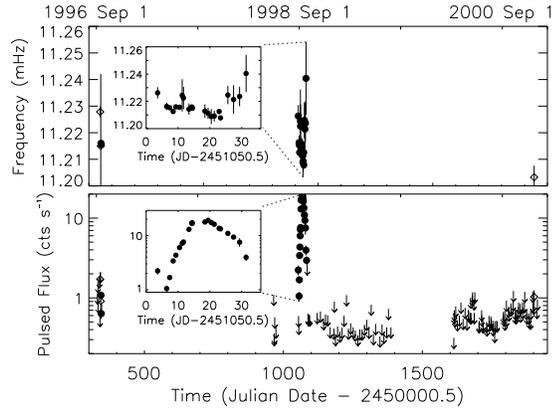}
\caption{(Top): Frequency search results for XTE J1906+09. Barycentric
corrections have been applied. Filled circles
indicate 99.99\% confidence detections and open diamonds indicate 99.9\%
confidence detections. The 99.9\% confidence detection on MJD 51902 is likely
due to contamination from 4U 1907+09. (Inset): Expanded view of pulse frequency
detections during the 1998 outburst.
(Bottom): Root mean squared 2-30 keV pulsed flux for XTE J1906+09. The count
rate in all observations has been corrected for the offset pointing direction
and scaled to the rate expected from 5 PCU if fewer PCU were active. Downward 
pointing arrows denote 99\% confidence upper limits. (Inset): Expanded view of 
pulsed fluxes corresponding to frequency detections during the 1998 outburst.
\label{fig:freq1906}}
\end{figure}

\begin{figure}
\plotone{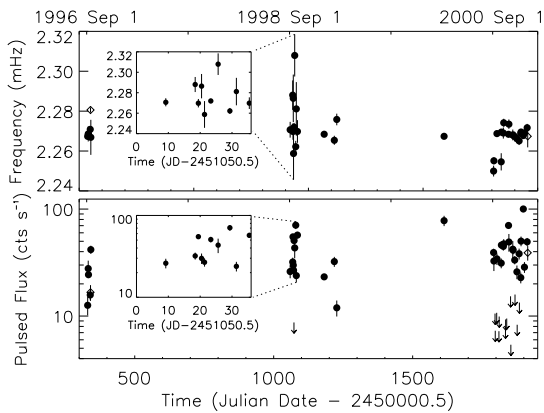} 
\caption{(Top): Frequency search results for 4U 1907+09. Barycentric
corrections have been applied. Filled circles
indicate 99.99\% confidence detections and open diamonds indicate 99.9\%
confidence detections. (Inset): Expanded view of 4U 1907+09 pulse frequency
detections during the 1998 outburst of XTE J1906+09. 
(Bottom): Root mean squared 2-30 keV pulsed flux for 4U 1907+09. The count
rate in all observations has been corrected for the offset pointing direction
and scaled to the rate expected from 5 PCU if fewer PCU were active. Downward 
pointing arrows denote 99\% confidence upper limits. (Inset): Expanded view of 
4U 1907+097 pulsed fluxes corresponding to 4U 1907+09 frequency detections
during the 1998 outburst of XTE J1906+09.
\label{fig:freq4u1907}}
\end{figure}

\begin{figure}
\plotone{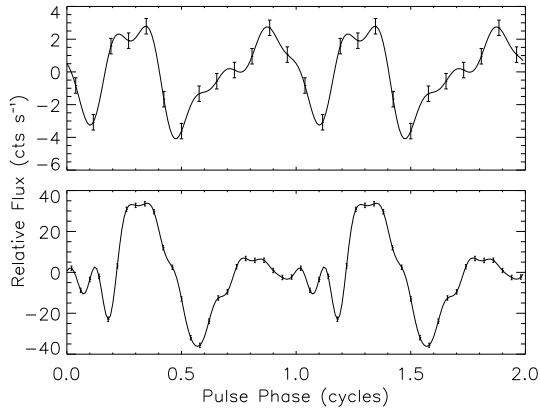}
\caption{(Top): Mean subtracted pulse profile from MJD 50343 (1996 September 17)
generated by fitting a Fourier expansion to 2-30 keV PCA event mode 
light curves. Only the first 6 harmonics were significant. (Bottom): Mean 
subtracted pulse profile from MJD 51072 (1998 September 16) generated by fitting
a harmonic Fourier expansion to 2-30 keV PCA Good Xenon data light curves. Only
the first 12 harmonics were significant. 
\label{fig:tamp}}
\end{figure}

\begin{figure}
\plotone{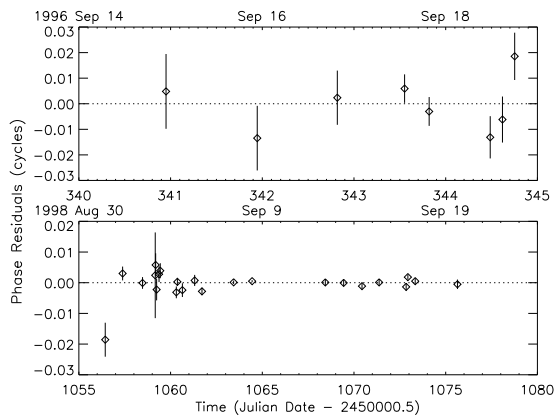}
\caption{(Top): Phase residuals for a fit of a third order polynomial to
connected pulse phase measurements from the 1996 outburst. (Bottom): Phase 
residuals for a fit to a fifth order polynomial to connected pulse phase 
measurements from the 1998 outburst.
\label{fig:resid}}
\end{figure}

\begin{figure}
\plotone{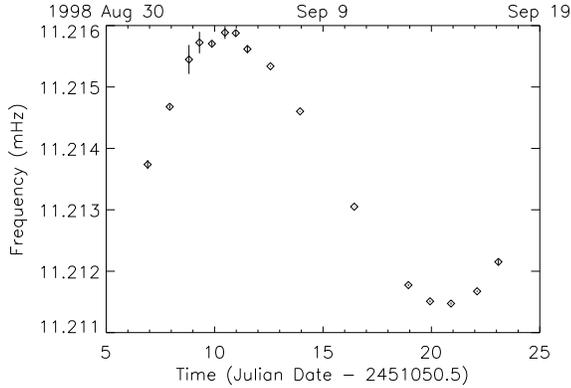}
\caption{(Top): Barycentered pulse frequencies computed from connected phases for the 1998 outburst.
A sinusoidal signature, likely due to orbital motion of the pulsar, is evident.
\label{fig:outburst2}}
\end{figure}

\begin{figure}
\plotone{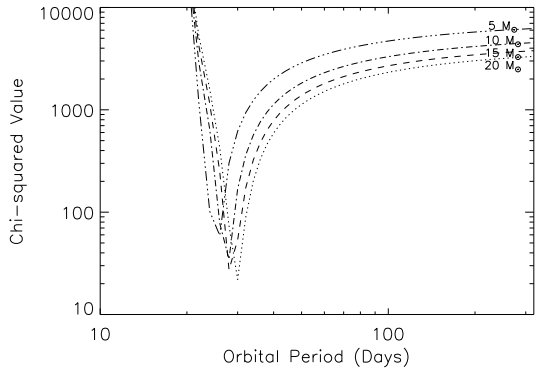}
\caption{$\chi^2$ versus orbital period for 4 mass functions $f(M)=5,10,15, 20
M_{\odot}$. The orbital period and mass function were fixed for each grid point.
The eccentricity, periapse angle, epoch of periastron passage, and the
parameters of the quadratic phase model were allowed to vary in each fit to 23
phase points, giving us 17 degrees of freedom.
\label{fig:porb_grid}}
\end{figure}

\begin{figure}
\plotone{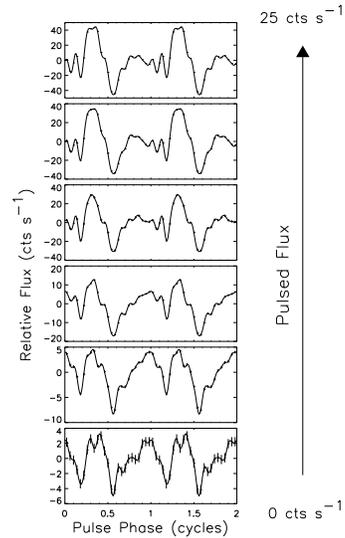}
\caption{(Top 5 panels): Mean subtracted pulse profiles from the 1998 outburst averaged over 
time in 5 bands of 2-30 keV rms pulsed flux: 0-5, 5-10, 10-15, 15-20, 20-25 
counts s$^{-1}$. The profiles were phase aligned using the polynomial model
in Table~\ref{tab:poly}. (Bottom panel): The pulsed profile from the brightest
interval in the 1996 outburst (also shown in Figure~\ref{fig:tamp}) which had 
an rms pulsed flux of $1.9 \pm 0.1$ counts s$^{-1}$. This profile has been aligned by
eye with the 1998 pulse profiles.
\label{fig:profile_intensity}}
\end{figure}

\begin{figure}
\plotone{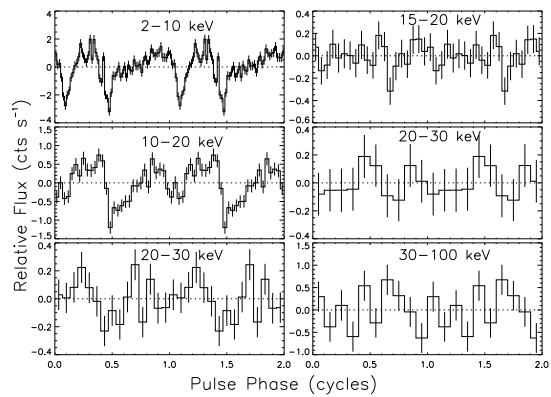}
\caption{Mean subtracted pulse profiles from PCA (left) and HEXTE (right) data 
from 1996 September 17-18 epoch-folded using the phase model in 
Table~\ref{tab:poly}. PCA profiles above 20 keV and all HEXTE profiles are 
consistent with a constant value. 
\label{fig:profile_energy1}}
\end{figure}

\begin{figure}
\plotone{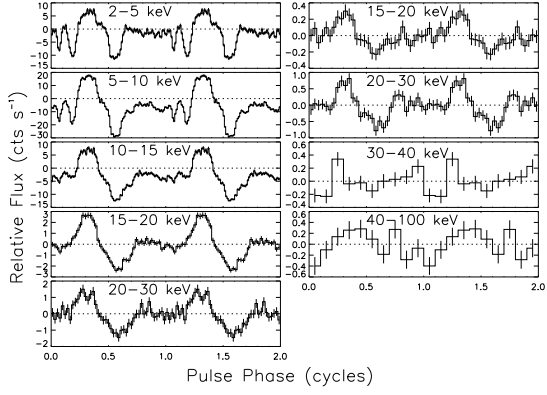}
\caption{Mean subtracted pulse profiles from PCA (left) and HEXTE (right) data 
from 1998 September 8 epoch-folded using the phase model in Table~\ref{tab:poly}. 
\label{fig:profile_energy2}}
\end{figure}

\begin{figure}
\plotone{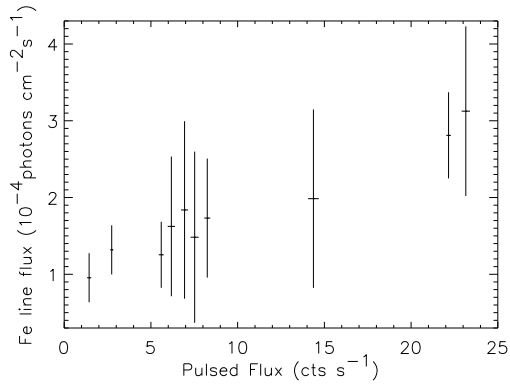}
\caption{Integrated flux in the iron line at $\sim 6.6$ keV versus 2-30 keV 
pulsed flux in counts s$^{-1}$. The iron line flux is correlated with
the pulsed flux, with a correlation coefficient of 0.93 and a chance probability
of $10^{-4}$, suggesting that the iron line is intrinsic to XTE J1906+09.
\label{fig:iron}}
\end{figure}

\begin{figure}
\plotone{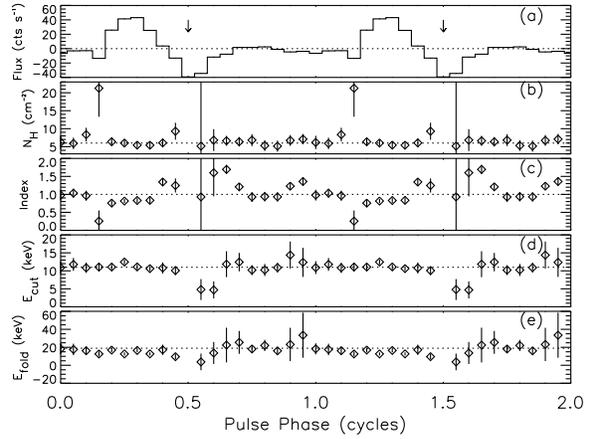}
\caption{Phase dependence of spectral parameters of an absorbed power law with 
a high energy cutoff. (a) The mean subtracted 2-30 keV pulse profile divided 
into 20 equal bins.  Each bin was fitted separately with XSPEC. The spectrum from pulse
minimum,  indicated using arrows on the plot, was used as the background
spectrum for all spectral fits. (b) The absorption column density $N_{\rm H}$
in units of  cm$^{-2}$. (c) The power law photon index. (d) The cutoff energy
$E_{\rm cut}$ in keV. (e) The folding energy $E_{\rm fold}$ in keV.
\label{fig:phaseresolved}}
\end{figure}

The two outbursts differ considerably in peak intensity and duration
(Figure~\ref{fig:freq1906}). The second outburst lasted about 25 days and was a
factor of $\sim 20$ brighter in 2-30 keV pulsed flux that the first one which
was very short, 2-4 days. Although 
the second outburst was brighter and longer, it is unlikely that it is a giant 
outburst like that observed in some Be X-ray binaries because giant outbursts are 
typically near Eddington luminosity at their peaks (e.g. Bildsten et al.\ 1997).
Assuming a distance of $\lesssim 20$ kpc, the 2-30 keV peak luminosity of the
second outburst is $\lesssim 2 \times 10^{37}$ ergs cm$^{-2}$ s$^{-1}$,
consistent with a normal outburst. In the pulse frequencies measured
by differencing connected phase measurements, we see a distinct sinusoidal 
signature that is likely due to the orbit of XTE J1906+09
(Figure~\ref{fig:outburst2}). Two possible scenarios could produce this 
signature: (1) a short ($\sim 30$ day), nearly circular, orbit or (2) periastron passage in a 
long, eccentric, orbit. Accreting X-ray pulsar systems with short, nearly 
circular, orbits are typically persistent systems \citep{Bildsten97} and XTE
J1906+09 is clearly a transient system. The second case of periastron passage in
a long, eccentric, orbit is what one would expect from a Be/X-ray transient 
system. Fits of a quadratic phase model and an orbit with a fixed orbital 
period and mass function yielded $\chi^2$ minima for orbital periods of 26-30
days and eccentricities of 0.03-0.06.  

An orbital period of 26-30 days is shorter than what would be expected from the
general correlation between orbital and pulse periods in Be/X-ray transients 
\citep{Corbet86, Waters89}. 4U 0728--25 \citep{Corbet97a} and
possibly GRO J2058+42 \citep{Wilson00, Wilson98, Corbet97b} may also have 
shorter orbital periods than expected. 4U 0728--25 is a confirmed Be/X-ray 
binary with a 103.2-s pulse period and a 34.5 day outburst period. GRO 
J2058+42 is a 195.6-s pulsar with either a 55 or 110 day orbital period. 
Odd-even pulsed flux variations in outbursts observed with BATSE may indicate 
the orbital period is 110 days \citep{Wilson98}; however, observations with the
{\em RXTE} ASM and PCA do not show clear evidence for differences between the 
outbursts, suggesting an orbital period of 55 days \citep{Wilson00, Corbet97b}.
In both cases, the periods are outburst recurrence times, so pulse timing analysis of 
data with appropriate phase coverage is needed to confirm these as orbital 
periods. These three sources may be members of an unusual class of Be/X-ray 
binary that does not lie on the general spin-orbital period correlation
\citep{Corbet86, Waters89} with most Be systems. Our grid searches also 
suggest that the eccentricity in XTE J1906+09 may be quite low. Four Be/X-ray 
binaries have measured orbits with low eccentricities, X Per ($e=0.11$, 
Delgado-Marti et al.\ (2001)), XTE J1543--568 ($e < 0.03$, in't Zand, Corbet,
\& Marshall 2001), GS 
0834--430 ($0.1 \lesssim e \lesssim 0.17$, Wilson et al.\ 1997), and 2S 
1553--542 ($e < 0.09$, Kelley, Rappaport, \& Ayasli 1983). The system that 
appears most similar to XTE J1906+09 is 2S 1553--542, which has a spin period of
9.3-s and an orbital period of 30.6 days. However, the orbit of this system was
measured using a 20 day observation and the source has not been observed in 
outburst again. In both 2S 1553--542 and XTE J1906+09, the orbital parameters 
were determined using less than one orbital period and should be treated with 
caution. Strong intrinsic torques, often correlated with the intensity of the 
pulsar \citep{Finger99, Bildsten97}, could easily produce misleading results in
a short stretch of data. Coupling between intrinsic torques and the orbit could
lead to erroneous orbital parameters. Observations of additional outbursts from
both sources are needed to confirm the orbital parameters. 

If XTE J1906+09 is a low eccentricity system, it is surprising that it appears
to show only isolated normal outbursts. At least one other system, Cep X-4
\citep{Wilson99} also showed isolated normal outbursts, but its orbit is
unknown. The flux level of 2S 1553--542 was not reported by \citet{Kelley83}, so
it is unclear whether it underwent a giant or normal outburst. The viscous 
decretion disk model, which accounts for most properties of Be stars, predicts 
the truncation of the circumstellar disk around the Be star, due to the presence
of the neutron star. In systems with low eccentricities ($e < 0.2$), this model
predicts that the truncation mechanism would be so effective that no normal 
outbursts would be seen at all. Only giant outbursts would be expected when the
disk was strongly disturbed \citep{Okazaki01}. However this model fails to 
explain the series of normal outbursts in GS 0834--430 \citep{Wilson97}. It also
does not easily explain the 103.2-s pulsar 4U 0728--26, which has not shown 
giant outbursts, but shows a 34.5 day modulation in its persistent flux 
\citep{Corbet97a}. Perhaps a moderate perturbation in the Be disk, not large 
enough to produce a giant outburst, could produce the isolated normal outbursts
we observe in XTE J1906+09.

The transient nature of XTE J1906+09 suggests that it is likely a member of a
Be/X-ray binary system. {\em RXTE} PCA scans during the second outburst of XTE
J1906+09 localized it to a 2\arcmin\ radius error circle centered on 
$\alpha = 19^h05^m20^s, \delta = 9\arcdeg 02\arcmin.5$, J2000 
\citep{Takeshima98b}. No objects were found in the error circle in the
Simbad\footnote{\url{http://simbad.u-strasbg.fr}} database. We searched the 
catalogs at the High Energy Astrophysics Science Archive Research 
Center\footnote{\url{http://heasarc.gsfc.nasa.gov}} for a possible counterpart.
The ROSAT All-Sky Survey Faint Source catalog contained no sources within 
30\arcmin of XTE J1906+09. Although pointed observations of SGR 1900+14 were 
made with EUVE and BeppoSAX, XTE J1906+09 was outside the field-of-view in all 
cases. The ASCA Galactic Plane Survey \citep{Sugizaki01} did not detect an X-ray
source brighter than $10^{-12.5}$ ergs cm$^{-2}$ s$^{-1}$ within 52\arcmin\ of 
XTE J1906+09. However, the 2\arcmin\ error circle contains no shortage of 
stars. In the USNO A-2 catalog\footnote{\url{http://www.nofs.navy.mil}}, we 
found 41 stars in the error circle, with B-magnitudes ranging from 15.9 to 20.0
and R-magnitudes ranging  from 14.0 to 18.4. If this sample of stars ranges over
all distances (as is likely in this case), each star will have its own color
correction depending on distance, so comparing objects in the field is
difficult. Our $N_{\rm H}$ values from Table~\ref{tab:spectra} imply an 
extinction of $A_V \gtrsim 15$ \citep{Predehl95}, which makes it unlikely we 
will see the companion in the visible range. One infrared source was found in 
the error circle at $\alpha=19^h05^m15.38^s, \delta=09\arcdeg 03\arcmin 39.6 
\arcsec$, J2000 in the MSX5C\footnote{\url{http://www.ipac.caltech.edu/ipac/msx/msx.html}} 
Infrared Point Source Catalog \citep{Egan99}; however, this object does not
appear in the Catalog of Infrared 
Observations\footnote{\url{http://ircatalog.gsfc.nasa.gov}}. An X-ray 
observation with a very sensitive imaging instrument such as {\em Chandra} or 
{\em XMM}, since XTE J1906+09 would most likely be in quiescence, is needed to 
improve the location accuracy before a companion can be identified. 

\acknowledgements
This research has made use of data obtained from the High Energy Astrophysics 
Science Archive Research Center (HEASARC), provided by NASA's Goddard Space 
Flight Center. This research has also made use of the SIMBAD database, operated
at CDS, Strasbourg, France. We thank M.J. Coe for helpful discussions.

\clearpage

\begin{deluxetable}{cccccc}
\tabletypesize{\scriptsize}
\tablecaption{{\em RXTE} PCA Observations with XTE J1906+09 in the PCA field of view}
\tablewidth{0pt}
\tablehead{
\colhead{Cycle} & \colhead{Proposal ID}   & 
\colhead{No. Obs.} & \colhead{On-Source Time (ksec)} &
\colhead{Dates} & \colhead{MJD} }
\startdata
AO1 & 10228 & 19 & 97.7 & 1996 Sep  4-1996 Sep 20 & 50330-50346 \\
AO3 & 30197 & 12 & 46.1 & 1998 May 31-1998 Aug 29 &  50964-51054 \\
AO3 & 30410 & 50 & 178.6 & 1998 May 29-1998 Dec 22 & 50962-51169 \\
AO4 & 40130 & 38 & 266.6 & 1999 Jan 3-1999 Jul 28 & 51181-51387 \\
AO4 & 40130 & 9 & 51.4 & 2000 Mar 10-2000 Mar 23 & 51613-51626 \\
AO5 & 50142 & 50 & 189.9 & 2000 Mar 29-2000 Aug 13 & 51632-51769 \\
AO5 & 50421 & 61 & 259.4 & 2000 Sep 4-2001 Jan 6  & 51791-51915 \\ 
\enddata
\tablecomments{Observation start and stop times can be obtained from the web-site
\url{http://heasarc.gsfc.nasa.gov}}
\label{tab:obs}
\end{deluxetable}

\clearpage
\begin{deluxetable}{ccccc}
\tabletypesize{\scriptsize}
\tablecaption{Percentage of time PCUs were active}
\tablewidth{0pt}
\tablehead{
\colhead{Proposal ID} & \colhead{5 PCUs on}   & 
\colhead{4 PCUs on} & \colhead{3 PCUs on} &
\colhead{2 PCUs on} }
\startdata
10228 & 58\% & 27\% & 15\% & 0\% \\
30197 & 8\% & 87\% & 5\% & 0\% \\
30410 & 58\% & 35\% & 6\% & 0\% \\
40130\tablenotemark{a} & 48\% & 39\% & 13\% & 0\% \\
40130\tablenotemark{b} & 0\% & 10\% & 90\% & 0\% \\
50142 & 2\% & 43\% & 52\% & 3\% \\
50421 & 16\% & 43\% & 39.5\% & 1.5\% \\
\enddata
\tablenotetext{a}{1999 observations}
\tablenotetext{b}{2000 observations}
\label{tab:pcus}
\end{deluxetable}

\clearpage
\begin{deluxetable}{lcc}
\tabletypesize{\scriptsize}
\tablecaption{Polynomial Fits to Pulse Phase Measurements}
\tablewidth{0pt}
\tablehead{
\colhead{Coefficient} & \colhead{Outburst 1}   & \colhead{Outburst 2}}
\startdata
$c_0$ (cycles) & $0.048 \pm 0.006$ & $-0.2938 \pm 0.0009$ \\
$c_1$ (cycles day$^{-1}$) & $969.008 \pm 0.008$ & $968.6979 \pm 0.0002$ \\
$c_2$ (cycles day$^{-2}$) & $(-6 \pm 1) \times 10^{-2} $ & $(-1.492 \pm 0.008)
\times 10^{-2}$ \\
$c_3$ (cycles day$^{-3}$) & $(-1.3 \pm 0.4) \times 10^{-2}$ & $(2.269 \pm 0.008) \times
10^{-3}$\\
$c_4$ (cycles day$^{-4}$) & \nodata & $(1.30 \pm 0.02) \times 10^{-4}$\\
$c_5$ (cycles day$^{-5}$) & \nodata & $(-2.8 \pm 0.1) \times 10^{-6}$\\
$t_0$ (MJD) & 50344.0 & 51068.906 \\
$\chi^2$/dof  & 9.7/4 & 39.0/17 \\
\enddata
\label{tab:poly}
\end{deluxetable}

\clearpage

\begin{deluxetable}{lll}
\tabletypesize{\scriptsize}
\tablecaption{Spectral Fitting Results}
\tablewidth{0pt}
\tablehead{
\colhead{Parameter} & \colhead{PCA model background} & \colhead{pulse minimum
background}}
\startdata
$N_{\rm H}$ & $(2.8 \pm 0.1) \times 10^{22}$ cm$^{-2}$ & $(6.0 \pm 1.0) \times
10^{22}$ cm$^{-2}$\\
Power Law Photon Index & $1.17 \pm 0.01$ & $1.01 \pm 0.08$ \\
Power Law Normalization & $(1.93 \pm 0.05) \times 10^{-2}$ & $(7 \pm 1) \times
10^{-3}$ \\
$E_{\rm line}$ & $6.69 \pm 0.05$ keV & \nodata \\
$\sigma_{\rm line}$ & $0.4 \pm 0.1$ keV  & \nodata \\
Gaussian Normalization & $(2.7 \pm 0.5) \times 10^{-4}$ & \nodata \\
$E_{\rm cutoff}$ & $13.3 \pm 0.2$ keV & $11 \pm 1$ keV \\
$E_{\rm fold}$ & $16.8 \pm 0.9$ keV & $19 \pm 4$ keV \\
Flux (2-10 keV) & $1.1 \times 10^{-10}$ ergs cm$^{-2}$ s$^{-1}$ & $7.2 \times
10^{-11}$ ergs cm$^{-2}$ s$^{-1}$ \\ 
$\chi^2/$dof & 57.96/67 &  66.98/70 \\
\enddata
\label{tab:spectra}
\end{deluxetable}

\clearpage

\end{document}